\documentclass[referee]{aa} 
\usepackage{amsmath}
\usepackage{graphicx}
\usepackage{txfonts}
\usepackage{subfigure}
\usepackage{natbib}
\bibpunct{(}{)}{;}{a}{}{,} 
\begin{document}
   \title{Analytical view of diffusive and convective cosmic ray transport in elliptical galaxies }
   \titlerunning{Cosmic ray transport in elliptical galaxies}

   \author{T. Hein
          \and
          F. Spanier
          }

   \offprints{F. Spanier, \\ \email{fspanier@astro.uni-wuerzburg.de}}

   \institute{Lehrstuhl f\"ur Astronomie, University of W\"urzburg,
             Am Hubland, D-97074 W\"urzburg\\
             }

   \date{Received 10.09.07, accepted 07.11.07}


  \abstract
   {An analytical solution of the generalized diffusive and convective transport
   equation is derived to explain the transport of cosmic ray protons within elliptical
   galaxies.}
   {Cosmic ray transport within elliptical galaxies is an interesting element in
   understanding
    the origin of high energetic particles measured on Earth. As
    probable sources of those high energetic particles, elliptical galaxies show a dense
    interstellar
    medium as a consequence of activity in the galactic nucleus or merging events between
    galaxies. Thus it is
    necessary for an appropriate description of cosmic ray transport
    to take the diffusive and convective processes in a dense interstellar environment into account.
    Here we show that the transport equations can be solved analytically
    with respect to the given geometry  and boundary conditions
    in position space,
    as well as in momentum space.}
   {From the relativistic Vlasov equation, which is the most fundamental equation for a
    kinetic description of charged particles within the interstellar medium in
    galaxies, one finds a
    generalized
    diffusion-convection equation in quasilinear theory. This has the form of a `leaky box'
    equation,
    meaning particles are able to escape the confinement region by diffusing out of the galaxy.
    We apply here the `diffusion approximation', meaning that diffusion in
    gyrophase and pitch angle are the fastest particle-wave interaction
    processes. An analytical solution can be obtained using the `scattering time
    method', i.e. separation of the spatial and momentum problems.
   }
   {The spatial solution is shown using a generalized source of cosmic rays. Additionally, the
    special case of a jet-like source is illustrated. We present the solution in momentum
    space with respect to an escape term for cosmic ray protons
    depending on the spatial shape of the galaxy. For a delta-shape injection
    function,
    the momentum solution is obtained analytically. We find that the spectral index measured
    on Earth can be obtained by appropriately choosing of the strength of Fermi I and Fermi II processes.
    From these results we calculate
    the gamma-ray flux from pion decay due to proton-proton
    interaction to give connection to observations. Additionally we determine the escape-spectrum of cosmic rays.
    The results show that both spectra are harder than the intrinsic power-law spectrum for cosmic
    rays in elliptical
    galaxies.
  }
   {}


   \maketitle
%

\section{Introduction}

  Since their discovery by Viktor Hess in 1912, cosmic rays have been one of the
  biggest fields of interest in astrophysics, and yet the origin of these
  particles is still an open question.
  Fully ionized atomic nuclei
  reach the Earth coming from outside the solar system with very high
  energies up to $10^{20}$ eV. Most of them with energies $< 10^{17}$ eV
  seem to originate in the
  Milky Way, while the highest energetic ones are considered to have an
  extragalactic origin \citep[for a review see][]{Hoer}.

  The most accepted model for the origin of ultrahigh
  energy cosmic rays (UHECRs) is acceleration in shock fronts due to
  Fermi processes \citep[]{Fermi49}; hence, the main sources are gamma ray
  bursts (GRBs), active galaxies
  like active galactic nuclei (AGN) \citep[]{Tav}, or colliding galaxies.
  The last two are specially interesting for two reasons.
  First UHECRs can be generated in elliptical galaxies. Second the increase
  in the interstellar medium density in objects of these
  types influences cosmic ray transport \citep[see][]{BekkiShioya}.
  But since all elliptical galaxies have an interstellar medium due to star winds, we conclude that
  the study of transport processes is interesting in general \citep[see][]{Knapp}.

  In particular,
  as a result of the GZK-effect, very close AGN are the most probable candidates
  for UHECR sources \citep[see][]{Bier}. One of these nearby AGN is the giant
  elliptical galaxy M87. It is proposed that this galaxy is responsible
  for acceleration of cosmic ray protons due to Fermi I processes \citep{Bland,Rieger} in
  shockfronts within the jet \citep{Reimer}. For an overview of proton acceleration in jets
  \citep[see][]{Mannheim}.
  Since particle acceleration in a jet is located within active
  galaxies surrounded
  by an interstellar medium, the high energetic
  protons undergo physical transport processes before they escape out of the galaxy and reach
  the detectors on Earth, making a model for cosmic ray transport
  in elliptical galaxies inevitable. This helps give an answer to the question about the
  origin of cosmic rays.

Progress has been made in the field of modeling cosmic ray transport
with the numerical description of transport processes by
\citet{OwJok} and \citet{StrMos}. Nevertheless we follow the basic
ideas presented in the underlying papers of \citet{LS85},
\citet{WangSchlick}, and \citet{LS88}, who used an analytical
description of cosmic ray transport. In relation to our work such, a
treatment has the following advantages: the model is adequate for
cosmic ray transport within any kind of elliptical galaxy including
arbitrary cosmic ray sources and the physical parameters involved in
our model can be easily fitted to measurements. After all, our
analytical model can serve as a test case for more profound
numerical models.

  In this paper we solve the cosmic ray transport equation analytically
  with respect to a kinetic description of
  the interstellar plasma in elliptical galaxies. Special attention is paid to the spatial
  transport of charged nuclei. In addition, the solution of the momentum equation is derived to explain
  general properties of this model. As a result, we present illustrative examples of spatial, as well as
  momentum, cosmic ray transport for given sources of charged nuclei.
  To show the connection to observations,
  we calculate the gamma-ray flux from neutral pion decay. These mesons are produced by inelastic
  scattering processes between cosmic ray protons. The resulting power-law spectrum is slightly harder
  than the intrinsic one for cosmic rays. Finally, we present the escape spectrum of charged particles
  leaving elliptical galaxies. Similar to the gamma-ray flux, this spectrum is flatter than the
  intrinsic one.


\section{Basic equations}

  To describe the propagation of cosmic ray nuclei within elliptical
  galaxies, we follow \citet{LS85}, \citet{LS88} and \citet{Schlick1}.
  For the description of transport processes they use the `diffusion
  approximation',
  which means that the fastest particle- plasma
  wave interaction processes are diffusion in gyrophase and pitch angle. Thus following
  \citet{Jokipii}, \citet{HassWib}, and \citet{Skilling}, we take an
  isotrope
  particle distribution function in momentum space. Here
  we idealise the interstellar medium
  as a homogeneous volume containing primary cosmic rays being accelerated from the thermal
  background medium and secondaries resulting from fragmentation of primaries having a negligible
  abundance in the background medium \citep[cf.][]{Haya, Cow}.

  The transport of these particles
  at large momenta ($p>10$ GeV $c^{-1}$ nucleus$^{-1}$) is described by the steady-state transport
  equation \citep[e.g.][]{Schlick83}.
  Such a treatment is suitable to short timescales of diffusive and convective
  processes compared to the dynamical timescale of the galaxy ($t_{\textrm{dyn}}\approx 10^9$ years).
  This is true in the case
  of high energetic particles. We assume a spatial diffusion coefficient K(\textbf{r}) of $10^{29}$
  cm$^2$
  s$^{-1}$ at $p=$1 GeV, which is slightly larger than the value measured in the Milky Way
  ($K(\textbf{r})=10^{28}$ cm$^2$, \citep[cf.][]{Schlick1} because of diffusive processes being less
  effective in elliptical galaxies. Consequently we get for protons with TeV-energy a timescale of
  $\approx 10^8$ years.
  Furthermore, the dynamical age of the galaxy has to be greater than
  the timescale of source variability to obtain an appropriate description. This is usually given, since
  the size of the accretion region onto the central black hole is of the order of a few light-days so that
  a maximal variability timescale of some days is assumed.

  At large momenta, spatial diffusion in turbulent magnetic fields
  dominates convection in the galactic wind, so that we find a transport equation for the
  phase space density $f(\textbf{r},p)$ in spatial coordinates $\textbf{r}$ and in the momentum
  coordinate $p$:
  \begin{equation}
  \mathcal{L}_{\textbf{r}}f+ \mathcal{L}_{p}f+S(\textbf{r},p)=0.
  \label{dgl1}
  \end{equation}
  The spatial operator $\mathcal{L}_{\textbf{r}}$ is defined by
  \begin{equation}
  \mathcal{L}_{\textbf{r}}(\textbf{r},p) \equiv \nabla \left[K(\textbf{r},p)\nabla \right]
  \label{spoper}
  \end{equation}
  containing spatial diffusion with the spatial diffusion coefficient
  $K(\textbf{r},p)=K(\textbf{r})\kappa(p)$, where $\kappa(p)$ denotes the dimensionless
  dependence on the momentum variable $p$ without loss of generality. The momentum
  operator
  \begin{equation}
  \mathcal{L}_{p}(\textbf{r},p) \equiv p^{-2} \frac{\partial}{\partial p} \left[
  p^2 D(p) \frac{\partial}{\partial p}-p^2\dot{p}_{\textrm{gain}}-p^2\dot{p}_{\textrm{loss}}\right]
  -\frac{1}{\tau_{c}}
  \label{momoper}
  \end{equation}
  describes momentum diffusion by second-order Fermi processes (D(p)), energy gain due to first order
  Fermi processes ($\dot{p}_\textrm{gain}$), as well as continuous ($\dot{p}_\textrm{loss}$) and catastrophic
 ($\tau_{c}$)
  momentum loss processes.
  As shown in \citet{LS85}, fully-ionized particles heavier than protons have the same Fermi acceleration
  rates as protons, so hereinafter momentum means momentum per nucleon.
  We are interested primarily in the behaviour of the cosmic ray primary
  spectrum, so we do not take secondary particles due to fragmentation of primaries into account.
  On the other hand, we allow fragmentation of primaries as a general loss process.
  \newline
  A link between spatial and momentum diffusion processes can be seen in the relation
  between the two diffusion coefficients
  \begin{equation}
  D(\textbf{r},p)=\frac{C_1 p^2}{K(\textbf{r},p)},
  \label{DiffCoeff}
  \end{equation}
  where  $C_1$ stands for the
  proportionality factor being independent of $\textbf{r}$ and $p$.
  This close connection arises from the same basic physical process behind
  spatial and momentum diffusion:
  Protons are scattered in pitch angle due to the magnetic fields of MHD plasma waves causing
  spatial diffusion along ordered magnetic field lines, whereas cyclotron damping of the electric field
  associated with MHD waves affects diffusion in momentum space.
  For that reason it is
  necessary to solve this model in spatial coordinates as well as in momentum coordinates to get an
  acceptable description of transport processes in elliptical galaxies.

  \section{`Scattering time' method}
  We use the `scattering time' method proposed by \citet{SunyTit}
  to get an important class
  of exact analytical solutions of Eq.(\ref{dgl1}) following \citet{WangSchlick}.
  This implies, that the spatial and momentum operators can be separated as
  \begin{equation}
  \mathcal{L}_{\textbf{r}}(\textbf{r},p)=h(p)\,\mathcal{O}_{\textbf{r}}
    (\textbf{r})
    \hspace{0.2cm},\hspace{0.2cm}
    \mathcal{L}_{p}(\textbf{r},p)=g(\textbf{r})\,\mathcal{O}_{p}(p) .
  \label{OpSep}
  \end{equation}
  For ease of exposition, the source function $S(\textbf{r},p)$ is also a product of two separable
  functions, i. e.,
  \begin{equation}
  S(\textbf{r},p)=q(\textbf{r})Q(p).
  \label{SourceSep}
  \end{equation}
  As an aside we note that the requirement Eq.(\ref{OpSep}) is trivially fulfilled,
  if $K(\textbf{r})$ is constant ($K(\textbf{r})=K_0$) and
  $\dot{p}_{\textrm{gain}}$, $\dot{p}_{\textrm{loss}}$, as well as $\tau_{e}$, are all independent of spatial variables.
  Thus we use the following model containing constant factors $a_n$ with $n=1,2,3,4$ for the proportionality factors
  independent of spatial
  and momentum coordinates.
  First we take $D(p)=a_2 p^2/\kappa(p)$ to describe the momentum diffusion coefficient
  solely in momentum space.
  Furthermore we assume
  \begin{equation}
  \dot{p}_{\textrm{gain}}=a_1 p /\kappa(p)
  \label{asspgain}
  \end{equation}
  telling that first-order Fermi acceleration is related to the (momentum dependend) spatial
  diffusion coefficient due to MHD plasma-wave-scattering interactions within the acceleration
  process. The continuous loss term $\dot{p}_{\textrm{loss}}$ is independent of spatial coordinates leading
  to
  \begin{equation}
  \dot{p}_{\textrm{loss}}(p)=-a_3 \rho (p).
  \label{assploss}
  \end{equation}
  Similarly, we set
  \begin{equation}
  \tau_{c} (p)=a_{4}^{-1}\theta (p)
  \label{asscloss}
  \end{equation}
  for the catastrophic loss time due to fragmentation.
  \newline
  Under these conditions, in addition to Eq.(\ref{OpSep}) and (\ref{SourceSep}), we can find the
  formal mathematical solution of Eq.(\ref{dgl1}) as a convolution
  of the spatial and momentum solution functions $T(\textbf{r})$ and
  $M(p)$ following \citet{dFP}:
  \begin{equation}
  f(\textbf{r},p)=\int_0^{\infty}du\,T(\textbf{r},u)M(p,u),
  \label{Convolution}
  \end{equation}
  where $T(\textbf{r},u)$ has to satisfy the given spatial boundary conditions, and
  \begin{equation}
  \frac{\partial T}{\partial u}=\frac{1}{g(\textbf{r})}\mathcal{O}_{\textbf{r}} T
  \label{Sp1}
  \end{equation}
  with
  \begin{equation}
  \mathcal{O}_{\textbf{r}}=\nabla \left[ K_0 \nabla \right] \equiv  K_0 \Delta
  \hspace{0.2cm},\hspace{0.2cm}
  g(\textbf{r})=1
  \label{SpOpSep}
  \end{equation}
  and the conditions
  \begin{equation}
  T(\textbf{r},u=0) = q(\textbf{r})/g(\textbf{r})=q(\textbf{r})
  \label{SpatialRand1}
  \end{equation}
  \begin{equation}
  T(\textbf{r},u=\infty) = 0.
  \label{SpatialRand2}
  \end{equation}
  Here, $M(p,u)$ has to satisfy the given spatial boundary conditions, and
  \begin{equation}
  \frac{\partial M}{\partial u}=\frac{1}{h(p)}\mathcal{O}_{p} M
  \label{Mom1}
  \end{equation}
  with
  \begin{equation}
  \mathcal{O}_p=\frac{1}{p^2} \frac{\partial}{\partial p} \left[\frac{a_2 p^4}{\kappa(p)}
  \frac{\partial}{\partial p}-\frac{a_1 p^3}{\kappa(p)}+a_3p^2 \rho(p)\right]-
  \frac{a_4}{\theta_4(p)}
  \hspace{0.2cm};\hspace{0.2cm}
  h(p)=\kappa(p)
  \label{MomOpSep}
  \end{equation}
  and the conditions
  \begin{equation}
  M(p,u=0) = Q(p)/h(p)=Q(p)/\kappa(p)
  \label{MomentumRand1}
  \end{equation}
  \begin{equation}
  M(p,u=\infty) = 0.
  \label{MomentumRand2}
  \end{equation}
  As a consequence of the formal mathematical solution, we note that we have to solve two
  partial differential Eqs.(\ref{Sp1}) and (\ref{Mom1}) instead of the much more complicated
  differential Eq.(\ref{dgl1}).

  \section{Results}
  \subsection{Spatial solution}
  The most convenient way to find the formal solution of Eqs.(\ref{Sp1}) and (\ref{Mom1})
  is to start with the spatial problem.
  As can be seen from Eq.(\ref{SpOpSep}), the spatial operator $\mathcal{O}_{\textbf{r}}$ is of
  Sturm-Liouville type \citep[cf.][]{Arfken} and therefore has a complete eigenfunction system
  $E_i(\textbf{r})$. As a consequence, the solution function $T(\textbf{r},u)$ can be expanded in this
  orthonormal system as
  \begin{equation}
  T(\textbf{r},u)=\sum_i A_i(\textbf{r})\, e^{-\lambda_i^2 u}.
  \label{EigExp}
  \end{equation}
  Here the $A_i(\textbf{r})$ are defined by
  \begin{equation}
  A_i(\textbf{r})=\alpha_i E_i(\textbf{r}),
  \label{ExpCoeffEigen}
  \end{equation}
  implying that the coefficients $\alpha_i$ weight each eigenfunction $E_i(\textbf{r})$. The
  $\lambda_i$ denote the eigenvalues of $\mathcal{O}_{\textbf{r}}$ implying the special spatial geometry.
  \newline
  \begin{figure}
   \centering
   \includegraphics[width=6.5cm]{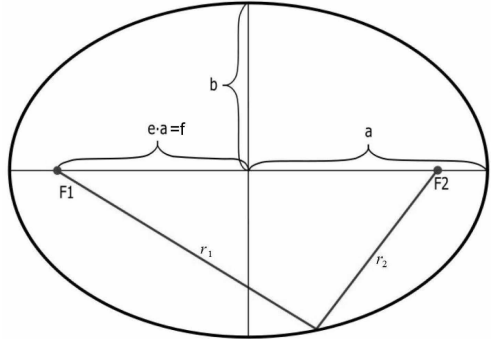}
      \caption{Schematical view of an ellipse with its fundamental properties.
              }
         \label{Ellipse1}
   \end{figure}
  The shape of elliptical galaxies is adjusted to the cosmic
  ray transport Eq.(\ref{SpOpSep}) using prolate spheroidal coordinates as they are defined by
  \citet{Abramowitz}:
  \begin{equation}
  \xi=\frac{r_1+r_2}{2f}
  \hspace{0.2cm},\hspace{0.2cm}
  \eta=\frac{r_1-r_2}{2f}
  \label{Defpsc}
  \end{equation}
  As can be seen from Fig.(\ref{Ellipse1}) $r_1$ and $r_2$ are the distances to the foci of the
  confocal ellipse, where $2f$ denotes the distance between the two foci $F_1$ and $F_2$. Additionally, we
  use the variable $\phi$ for the usual azimuthal dependence like in spherical coordinates. The
  following relations give the relation between these coordinates and the semi-major axis $a$ and
  the semi-minor axis $b$, respectively:
  \begin{equation}
  a=f\xi
  \hspace{0.2cm},\hspace{0.2cm}
  b=f\sqrt{\xi^2-1}.
  \label{SemiAxes}
  \end{equation}
  The numerical excentricity $e$ has a direct relationship to the coordinate $\xi$ via
\begin{equation}
e \equiv f/a=1/\xi.
\label{Ellip}
\end{equation}
  The variable $\eta$ is defined as $\eta=\cos\,\theta$ with $\theta$ the
  angle between the line on which the foci lie, and an arbitrary point on the ellipse.
  As a result the variables are defined in the range
  \begin{equation}
   \xi\, \in\,[1;\infty[
   \hspace{0.2cm},\hspace{0.2cm}
   \eta\, \in\,[-1;1]
   \hspace{0.2cm},\hspace{0.2cm}
   \phi\, \in\,[0;2\pi].
   \label{Varrange}
  \end{equation}
\begin{figure}[h!!!!!!]
 \centering
   \includegraphics[width=8cm]{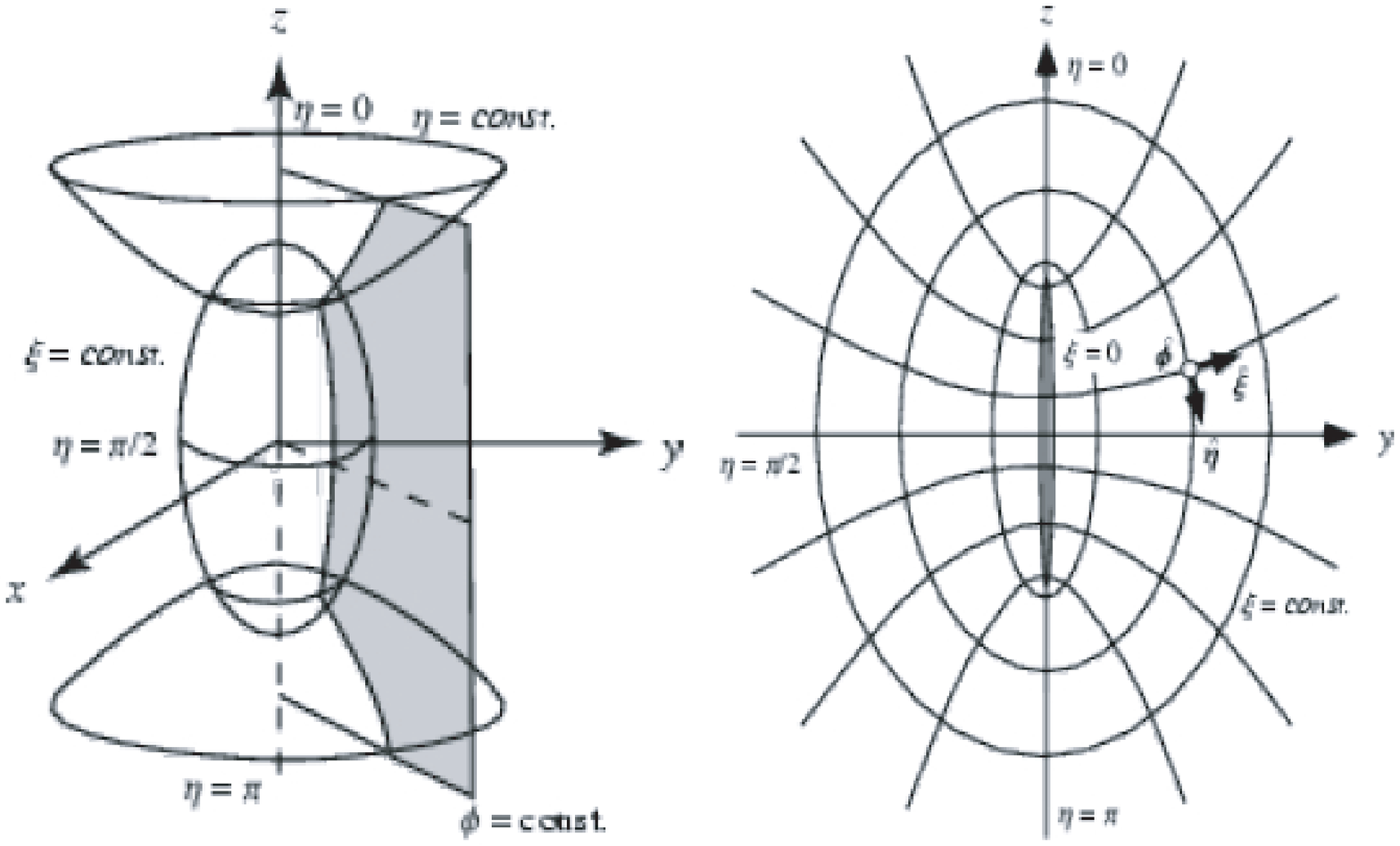}
\caption{\textit{Left:} Gray-shaded plane cuts ellipse leading to a
cut view like the graph on the left-hand
 side.
\newline
\textit{Right:} Direction of unit vectors of the variables $\xi$,
$\eta$, and $\phi$ in prolate spheroidal coordinates.
\newline
From \citet{Wolfram}.} \label{PSCWolfram}
\end{figure}

  Figure (\ref{PSCWolfram}) shows an illustration of the definition of the three spatial variables
  $\xi$, $\eta$, and $\phi$.
Because of these definitions, we can write Eq.(\ref{SpOpSep}) as
\begin{eqnarray}
\frac{\partial{T}}{\partial{u}}=
\frac{K_0}{f^2(\xi^2-\eta^2)} &\times & \left\{\frac{\partial}{\partial
\xi}\left[(\xi^2-1)\frac{\partial T}{\partial \xi}\right]+
\frac{\partial}{\partial \eta}\left[(1-\eta^2)\frac{\partial T}{\partial
\eta}\right]+
\right.
\nonumber \\
& &\,\left.+\frac{\xi^2-\eta^2}{(\xi^2-1)(1-\eta^2)}\frac{\partial^2
T}{\partial \phi^2}\right\} \label{DGL}.
\end{eqnarray}
Here we used the Laplacian in prolate spheroidal coordinates. The
general solution can be obtained by consecutive separation of
variables (see Appendix A):
\begin{eqnarray}
T(\xi,\eta,\phi, u)&=&\sum_{m,n} R_{mn}^{(1)}(c,\xi)\times S_{mn}^{(1)}(c,\eta)\times
\cos(m\phi)\times \exp(-k^2u)=
\nonumber \\
&=&\sum_{m,n} \left\{\left[\sum_{r=0,1}^{\infty ^{\prime}}
\frac{(2m+r)!}{r!}d_r^{mn}\right]^{-1}\times \left(\frac{\xi^2-1}{\xi^2}\right)^{m/2} \times\right.
\nonumber \\
& &\left.\times \sum_{r=0,1}^{\infty ^{\prime}}
\frac{(2m+r)!}{r!}d_r^{mn} \sqrt{\frac{\pi}{2c\xi}}J_{n+\frac{1}{2}}(c\xi) \times\right.
\nonumber \\
& &\left. \times\hat{d}_r^{mn}(c)P_{m+r}^{m}(\eta) \times
\cos(m\phi)\times \exp(-k^2u) \right\}.
\label{GenSol}
\end{eqnarray}
Here the $P_{m+r}^{m}(\eta)$ denotes associated Legendre functions
of the first kind, order $m+r$, and the $J_{n+\frac{1}{2}}(c\xi)$
denote Bessel functions of the first kind and of order
$n+\frac{1}{2}$. The sum  is extended over even values of $r$ as
indicated by the mark $^{\prime}$. The factor $c$ is given by
$c=\frac{fk}{\sqrt{K_0}}$.

To define reasonable boundary conditions, we assume a `leaky box'
model. Cosmic ray particles are trapped by disordered magnetic
fields within the confinement region of an elliptical galaxy. In
this they undergo diffusive and convective movements. At the edge of
the box, leakage out of the confinement area is possible.

As an illustrative example for spatial boundary conditions, we show
the solution depending on a constant source function over the
elliptical galaxy in Appendix B. To be more specific, we take a
jet-like source function here. The jet points in the direction
$\eta_{\textrm{inj}}$ (represented by a Dirac delta function) with a
length scale chosen to be $f_{\textrm{max}}$ for any choice of $\xi$
being smaller than an arbitrary maximum value of the confinement
region $\xi_c$. Particles leak out at the edge of this region. Such
a boundary condition is known in the literature as a `free-escape'
condition.  For a realistic assumption we decide to let the jet end
smoothly (see the `Fermi' function in Eq.(\ref{BoundUjet})). We
neglect any dependence on $\phi$ for an adequate illustration. These
conditions are taken into account by
\begin{equation}
T(\xi=\xi_{c},\eta,\phi,u)=0, \label{BoundXijet}
\end{equation}
by a periodical boundary condition in $\eta$
\begin{equation}
T(\xi,\eta = \-1,\phi,u)=T(\xi,\eta = +1,\phi,u),
\label{BoundEtajet}
\end{equation}
and by
\begin{equation}
T(\xi,\eta,u=0)=q_0\frac{\delta(\xi-\xi_c)\delta(\eta_{\textrm{inj}}-\eta)}
{\left[\exp(4(f-f_{\textrm{max}})+1\right]}.\label{BoundUjet}
\end{equation}
Finally after some calculations in which Eq.(\ref{GenSol}) has to
match the boundary conditions Eq.(\ref{BoundXijet}-\ref{BoundUjet}),
we derive the general solution (cf. Appendix B):
\begin{equation}
E_{i,r}(\textbf{r}) \equiv
T(\xi,u)=\sum_{i=1}^{\infty}\sum_{r=0}^{\infty ^{\prime}}\alpha_{ir}
\frac{1}{\sqrt{\xi}}J_{r+\frac{1}{2}}\left(y_{ir}\frac{\xi}{\xi_{c}}\right)P_r^0(\eta)
\exp \left[ -\frac{K_0 y_{ir}^2}{\xi_c^2 f^2}u \right ].
\label{TSolutionjet}
\end{equation}
The weighting factors are
\begin{equation}
\alpha_{ir}=\frac{q_0
P_r^0(\eta_{\textrm{inj}})}{\exp\left[4(f-f_{\textrm{max}})\right]+1}\times
\frac{\int_1^{\xi_c}\xi^{3/2}J_{r+\frac{1}{2}}\left(y_{ir}\frac{\xi}{\xi_{c}}\right)
d\xi} {\frac{\xi_c^2}{2}\left[ J_{r+\frac{3}{2}}(y_{ir})\right]^2}.
\label{diTildeXiLjet}
\end{equation}
Here the $y_{ir}$ stands for zeros of $J_{r+\frac{1}{2}}(x)$. The
sum over $r$ is extended over even values of this parameter. We see
that the general solution of Eq.(\ref{DGL}) with respect to spatial
boundary conditions indeed has the form of the eigenfunction
expansion Eq.(\ref{EigExp}) with Eq.(\ref{ExpCoeffEigen}).
 \begin{figure}
  \includegraphics[width=12cm]{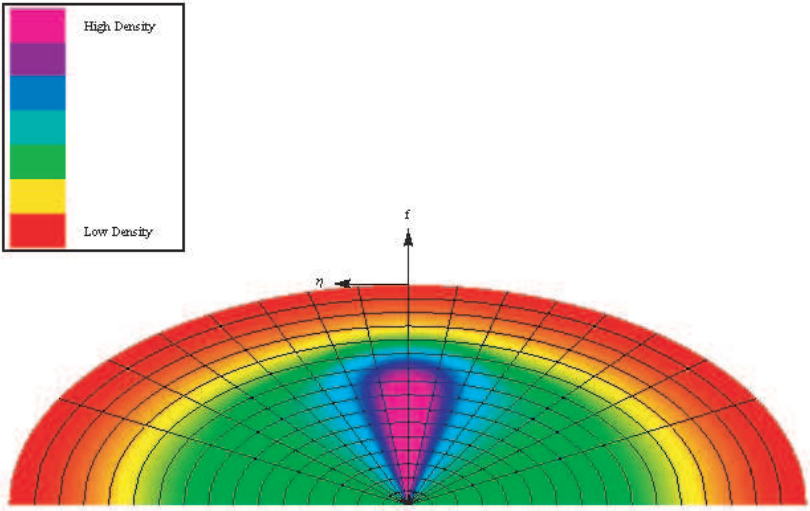}
     \caption{Graphic demonstration of the cosmic ray particle density given by the solution
Eq.(\ref{TSolutionjet}) with the weighting factors
(\ref{diTildeXiLjet}). The cosmic ray particle density is normalised
and given in arbitrary units. As an effect of chosen coordinates
only `one half' of the galaxy is visible. For illustration we
plotted $T(f,\eta,u=0)$ with a constant $\xi \equiv 1.2$ specifying
an E4 elliptical galaxy.
              }
        \label{fig:SpSol}
   \end{figure}
This solution is shown in Fig. (\ref{fig:SpSol}) with respect to
$100$ spatial eigenfunctions ($i_{\textrm{max}}=10$,
$r_{\textrm{max}}=18$). The jet is responsible for the cosmic ray
particles distributed over the whole galaxy due to diffusive
processes. It is broadened with a larger distance to the centre.
This effect is caused by the chosen geometry, whereas the loss of
magnetic collimation in real astrophysical sources can provide this.
To illustrate the shape of an elliptical galaxy we plotted
$T(f,\eta,u=0)$ with a constant $\xi \equiv 1.2$ specifying an E4-
galaxy. Using a more complicated source function, a better physical
description of particle distribution within elliptical galaxies can
be obtained.
\subsection{Consistency checks of the formal spatial solution}
To get a better understanding of our model, we prove the formal
mathematical solution (Eq.(\ref{GenSol})) of the spatial cosmic ray
transport equation. Our discussion is related to the solution found
in the illustrative example in Appendix B, but can similarly done
with Eq.(\ref{GenSol}):
\begin{equation}
T(\xi,u)=\sum_{i=1}^{\infty}\alpha_i
\frac{1}{\sqrt{\xi}}J_{\frac{1}{2}}\left(y_i\frac{\xi}{\xi_{c}}\right)
\exp \left[ -\frac{K_0 y_i^2}{\xi_c^2 f^2}u \right ]
\label{TSolutionCC}
\end{equation}
with the weighting factors
\begin{equation}
\alpha_i=\frac{q_0
\int_1^{\xi_g}\xi^{3/2}J_{\frac{1}{2}}\left(y_i\frac{\xi}{\xi_{c}}\right)
d\xi} {\frac{\xi_c^2}{2}\left[ J_{\frac{3}{2}}(y_i)\right ]^2}.
\label{diTildeXiLCC}
\end{equation}
Here we used periodical boundary conditions for $\eta$ and $\phi$
\begin{equation}
T(\xi,\eta = -1,\phi,u)=T(\xi,\eta = +1,\phi,u), \label{BoundEtaCC}
\end{equation}
\begin{equation}
T(\xi,\eta,\phi=2\pi,u)=T(\xi,\eta,\phi=0,u), \label{BoundPhiCC}
\end{equation}
and
\begin{equation}
T(\xi,\eta,\phi,u=0)=q_0
\Theta(\xi_{g}-\xi).\label{BoundUCC}
\end{equation}
\newline
First, this solution has to accomplish the given spatial boundary
conditions. As suggested in Eq.(\ref{BoundUCC}) we used a constant
source function over the whole size of the galaxy in order to lose
all angular dependencies. This behaviour is caused by the source
function not having any dependence in $\eta$ and $\phi$, but also
due to the periodic boundary conditions chosen for these variables.
Such an effect can also be seen in a spherical or disc geometry. To
the spherical geometry as performed by \citet{Sievers}, we added
angular dependencies and a constant source function over the galaxy
as an example. The weighting factors of the general solution
functions turn out to disappear in any case except for the angular
separation constants being zero (the angular solution functions are
spherical harmonics in this case), which means that there is no
dependence on these variables as it is expected.
\begin{figure}[h!]
   \centering
   \includegraphics[width=8cm]{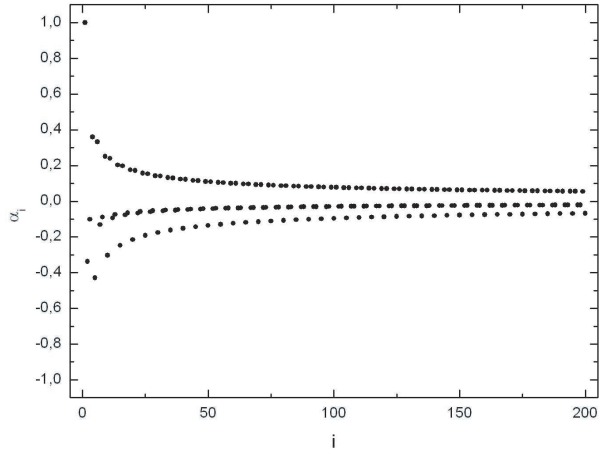}
      \caption{Normalised weighting factors $\alpha_i$ dependent on the number of zeros $y_i$ in
              Eq.(\ref{TSolutionCC})}
         \label{fig:ExpCoeff}
   \end{figure}

Second, we show that the sum of solution functions (cf.
Eq.(\ref{GenSol})), together with the weighting factors in addition
to the given spatial boundary conditions converges. For the
illustrative example in Appendix B, the normalised expansion
coefficients $\alpha_i$  are shown in Fig.(\ref{fig:ExpCoeff}). To
obtain applicable results it is essential to include only the first
few ones depending on the requested accuracy.
\begin{figure}[h!]
   \centering
   \includegraphics[width=8cm]{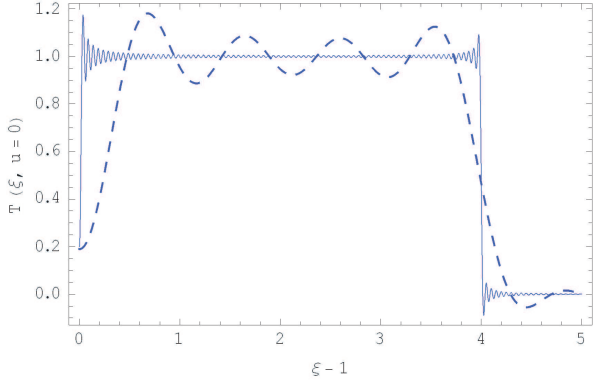}
      \caption{Spatial solution Eq.(\ref{TSolutionCC}) in addition with weighting factors
               $\alpha_i$ (Eq.(\ref{ExpCoeffEigen})) as a function of $\xi$. As an illustration, we put
               $\xi_g=4$. For the solution represented by the solid line we included $200$
               eigenvalues, whereas in the dashed solution only $10$ eigenvalues are taken into account.}
         \label{fig:SpatSolVergl}
   \end{figure}
The resulting solution function is shown in
Fig.(\ref{fig:SpatSolVergl}). In the special case of a constant
source function, the solution reproduces the boundary condition
$\theta(\xi_g-\xi)$ for the variable $\xi$. We used $\xi_g=4$ for
illustration. If we take many eigenfunctions into account, we see at
the discontinuity points $\xi=0$ and $\xi=\xi_g=4$ an oscillatory
phenomenon (Gibbs phenomenon). But in this case a closer solution
for the given boundary condition is obtained.

In analytical calculations, it is a common assumption to include
only the first spatial eigenvalue $\lambda_1$. This one is
associated with the longest escape timescale being the most
important one for modelling escape of particles out of the galaxy.
For numerical purposes, the computing time gives an upper limit to
the possible number of eigenvalues. Furthermore, the spatial
solution as performed in this paper has to match the formal solution
for a spherical geometry within the limit of small ellipticity ($e
\rightarrow 0$). \citet{Sievers} found as the spatial solution
\begin{equation}
T(R,u)=\sum_{m=1}^{\infty}c_m \frac{1}{\sqrt{R}}J_{\frac{1}{2}}\left( y_m \frac{R}{R_1}\right)
\exp\left(-\frac{K_0^2y_m^2}{R_1^2}u \right),
\label{SieversSolution}
\end{equation}
which already shows affinity to our solution Eq.(\ref{TSolutionCC}).
The $y_m$ denotes the zeros of $J_{\frac{1}{2}}(x)$ (here we
corrected \citet{Sievers}) and $R_1$ describes the edge of the
galaxy. Generally we can write for both solutions:
\begin{equation}
T(X,u)=\sum_{i=1}^{\infty} \alpha_i Z_i(c_iX)\exp(-\lambda_i^2u)
\label{generalisedSolution}
\end{equation}
The asymptotic behaviour of the solution Eq.(\ref{TSolutionCC}) is
with $Z(c_iX)\equiv \xi^{-1/2} J_{\frac{1}{2}}(c_i\xi)$
\begin{equation}
Z(c_i\xi)\xrightarrow{c_i\xi\rightarrow \infty} \frac{1}{c_i\xi}\cos
\left(c_i\xi-\frac{1}{2}\pi \right). \label{AsymEllipSolXi}
\end{equation}
The limit $c_i\xi\rightarrow \infty$ satisfies a spherical symmetry
with a numerical excentricity equal to zero (cf. Eq.(\ref{Ellip})).
Furthermore with this limit, the semi-major axis $c_i\xi$ goes to
infinity. Performing the same limit to Eq.(\ref{SieversSolution})
with $Z(c_iX)\equiv R^{-1/2}J_{\frac{1}{2}}(c_iR)$, we get
\begin{equation}
Z(c_iR)\xrightarrow{c_iR\rightarrow \infty} \frac{1}{c_iR}\cos \left(c_iR-\frac{1}{2}\pi \right),
\label{AsymEllipSolR}
\end{equation}
which is equal to Eq.(\ref{AsymEllipSolXi}). As a consequence we
showed that our solution functions embody the spherical geometry
within the limit of numerical excentricity equal to zero. It is also
straightforward to show that the weighting factors have the same
structure as those in \citet{Sievers} so that the general solution
is correct.

\subsection{Momentum solution}
\label{MomentumSolution} For the formal momentum solution, it is
necessary to have a closer look at the spatial solution. As noted
above, the eigenfunction expansion Eq.(\ref{EigExp}), in addition to
Eq.(\ref{ExpCoeffEigen}), is indeed the best way to solve the
spatial transport Eq.(\ref{Sp1}). Inserting this expansion into the
convolution Eq.(\ref{Convolution}), we can write
\begin{equation}
f(\textbf{r},p)=\sum_i \,A_i(\textbf{r})R_i(p)
\label{ConvAfterSpSol}
\end{equation}
with
\begin{equation}
R_i(p) \equiv \int_0^{\infty} du \,M(p,u)e^{-\lambda_i^2u}.
\label{RiDef}
\end{equation}
Consequently the momentum solution $R_i(p)$ obeys the ordinary
differential equation
\begin{equation}
\mathcal{O}_p R_i(p) - \lambda_i^2 g(p)R_i(p)=-Q(p),
\label{SpLeakyBox}
\end{equation}
which can be seen after multiplying Eq.(\ref{Mom1}) by
$T(\textbf{r},u)$ as given in Eq.(\ref{EigExp}) and integrating over
the convolution variable $u$ from $0$ to $\infty$. Each spatial
eigenvalue $\lambda_i^2$ enters this equation in the form of an
inverse catastrophic loss time for particle escape out of the
galaxy. Therefore Eq.(\ref{SpLeakyBox}) is called the `leaky box'
equation in momentum space.

As the result of the eigenfunction Eq.(\ref{EigExp}), we have to
solve one ordinary differential equation for each spatial eigenvalue
instead of the partial differential Eq.(\ref{Mom1}). Following
\citet{LS88}, we introduce some simplifying assumptions in order to
find  analytic solutions of Eq.(\ref{SpLeakyBox}). We take in
Eq.(\ref{MomOpSep}) $\theta_4=1$, $\rho(p)=p$, and
$\kappa(p)=(p/p_1)^{s}$ where $p_1$ is a normalisation value. These
assumptions imply that the most important continuous loss process in
elliptical galaxies at energies $>10$ GeV is adiabatic energy loss
due to a high galactic wind gradient, whereas pion production losses
are neglected because of the low number density of HI and HII. The
fragmentation lifetime ($\propto \theta_4$) is independent of
momentum, and the momentum dependence of the diffusion coefficient
is defined by the parameter $s=2-q$, where $q$ is the spectral index
of the magnetic turbulence power spectrum. Therefore we get
\begin{eqnarray}
&&\frac{p_{1}^{s}}{p^{2+s}} \frac{\partial}{\partial p} \left[ a_{2} p_{1}^{s} p^{4-s}\frac{\partial R_{i}}{\partial p} -a_{1} p_{1}^{s} p^{3-s} R_{i} + a_{3} p^{3} R_{i} \right]-
\nonumber \\
&&-\left[ \frac{a_{4} p_{1}^{s}}{p^{s}} +\lambda_{i}^{2} \right]
R_{i}=-Q(p)p^{-s} p_{1}^{s}. \label{MomProblemEigen}
\end{eqnarray}
This ordinary differential equation can be solved by a standard
technique taking the finiteness of $R_i$ at $p\rightarrow 0$ and
$p\rightarrow\infty$ into account. The formal mathematical solution
is taken from \citet{LS88}. However, we found after correction of
some minor mistakes:
\begin{eqnarray}
R_i(p)&=&
(a_2p_1^s)^{-1}\frac{\Gamma\left[\frac{a+3}{2s}+\frac{\xi_1-[(3+a)/2]\beta_1}{s(\beta_1^2+\psi_i)
^{1/2}} \right]}{\Gamma\left[\frac{a+3}{s} \right]} \times \left[\frac{\beta_1^2+4\psi_i}{s^2}\right]
^{\frac{a+3}{2s}} \times
\nonumber \\
&&\times p^a \exp\left[-\frac{\beta_1+(\beta_1^2+4\psi_i)^{1/2}}{2s}p^s \right]\times
\nonumber \\
&&
\times\left\{U\left[\frac{a+3}{2s}+\frac{\xi_1-[(3+a)/2]\beta_1}{s(\beta_1^2+4\psi_i)
^{1/2}},\frac{a+3}{s},\frac{(\beta_1^2+4\psi_i)^{1/2}}{s}p^s\right] \times \right.
\nonumber \\
&&\left.\times\int_0^pdp_0\,p_0^{s+1}Q(p_0)\exp\left[\frac{\beta_1-(\beta_1^2+4\psi_i)^{1/2}}{2s}
p_0^s\right]\times \right.
\nonumber\\
&&\left. \times M\left[\frac{a+3}{2s}+\frac{\xi_1-[(3+a)/2]\beta_1}{s(\beta_1^2+4\psi_i)
^{1/2}},\frac{a+3}{s},\frac{(\beta_1^2+4\psi_i)^{1/2}}{s}p_0^s\right]+\right.
\nonumber \\
&&\left.+M\left[\frac{a+3}{2s}+\frac{\xi_1-[(3+a)/2]\beta_1}{s(\beta_1^2+4\psi_i)
^{1/2}},\frac{a+3}{s},\frac{(\beta_1^2+4\psi_i)^{1/2}}{s}p^s\right] \times \right.
\nonumber \\
&&\left.\times\int_p^{\infty}dp_0\,p_0^{s+1}Q(p_0)\exp\left[\frac{\beta_1-(\beta_1^2+4\psi_i)^{1/2}}{2s}
p_0^s\right]\times \right.
\nonumber\\
&&\left. \times
U\left[\frac{a+3}{2s}+\frac{\xi_1-[(3+a)/2]\beta_1}{s(\beta_1^2+4\psi_i)
^{1/2}},\frac{a+3}{s},\frac{(\beta_1^2+4\psi_i)^{1/2}}{s}p_0^s\right]\right\}.
\nonumber \\
\label{MomSolGeneral}
\end{eqnarray}
Here, we took for brevity
\begin{equation}
\beta_1 \equiv a_3/(a_2p_1^s), \label{Beta1}
\end{equation}
\begin{equation}
a \equiv a_1/a_2, \label{Aa}
\end{equation}
\begin{equation}
\xi_1\equiv a_4/(a_2p_1^s), \label{Xi1}
\end{equation}
and
\begin{equation}
\psi_i\equiv \lambda_i^2/(a_2 p_1^{2s})
\label{psi}
\end{equation}
The functions $U$ and $M$ denote confluent hypergeometric functions
of first (Kummer) and second (Whittaker) order, respectively.
\newline
For the special case of a $\delta$-function injection of cosmic ray particles,
\begin{equation}
Q(p)=\delta(p-p_{\textrm{inj}}),
\label{DeltaInjection}
\end{equation} the solution Eq.(\ref{MomSolGeneral}) can be evaluated analytically. Therefore
we get for $p<p_{\textrm{inj}}$
\begin{eqnarray}
R_i(p)&=&
(a_2p_1^s)^{-1}\frac{\Gamma\left[\frac{a+3}{2s}+\frac{\xi_1-[(3+a)/2]\beta_1}{s(\beta_1^2+\psi_i)
^{1/2}} \right]}{\Gamma\left[\frac{a+3}{s} \right]} \times \left[\frac{\beta_1^2+4\psi_i}{s^2}\right]
^{\frac{a+3}{2s}} \times
\nonumber \\
&&\times p^a \exp\left[-\frac{\beta_1+(\beta_1^2+4\psi_i)^{1/2}}{2s}p^s \right]\times
\nonumber \\
&&
\times\left\{M\left[\frac{a+3}{2s}+\frac{\xi_1-[(3+a)/2]\beta_1}{s(\beta_1^2+4\psi_i)
^{1/2}},\frac{a+3}{s},\frac{(\beta_1^2+4\psi_i)^{1/2}}{s}p^s\right] \times \right.
\nonumber \\
&&\left.\times p_{\textrm{inj}}^{s+1}\exp\left[\frac{\beta_1+(\beta_1^2+4\psi_i)^{1/2}}{2s}
p_{\textrm{inj}}^s\right]\times \right.
\nonumber\\
&&\left. \times
U\left[\frac{a+3}{2s}+\frac{\xi_1-[(3+a)/2]\beta_1}{s(\beta_1^2+4\psi_i)
^{1/2}},\frac{a+3}{s},\frac{(\beta_1^2+4\psi_i)^{1/2}}{s}p_{\textrm{inj}}^s\right]\right\},
\nonumber \\
\label{MomSolDeltasmallerpi}
\end{eqnarray}
and for $p>p_{\textrm{inj}}$
\begin{eqnarray}
R_i(p)&=&
\times(a_2p_1^s)^{-1}\frac{\Gamma\left[\frac{a+3}{2s}+\frac{\xi_1-[(3+a)/2]\beta_1}{s(\beta_1^2+\psi_i)
^{1/2}} \right]}{\Gamma\left[\frac{a+3}{s} \right]} \times \left[\frac{\beta_1^2+4\psi_i}{s^2}\right]
^{\frac{a+3}{2s}} \times
\nonumber \\
&&\times p^a \exp\left[-\frac{\beta_1+(\beta_1^2+4\psi_i)^{1/2}}{2s}p^s \right]\times
\nonumber \\
&&
\times\left\{U\left[\frac{a+3}{2s}+\frac{\xi_1-[(3+a)/2]\beta_1}{s(\beta_1^2+4\psi_i)
^{1/2}},\frac{a+3}{s},\frac{(\beta_1^2+4\psi_i)^{1/2}}{s}p^s\right] \times \right.
\nonumber \\
&&\left.\times p_{\textrm{inj}}^{s+1}\exp\left[\frac{\beta_1+(\beta_1^2+4\psi_i)^{1/2}}{2s}
p_{\textrm{inj}}^s\right]\times \right.
\nonumber\\
&&\left.\times M\left[\frac{a+3}{2s}+\frac{\xi_1-[(3+a)/2]\beta_1}{s(\beta_1^2+4\psi_i)
^{1/2}},\frac{a+3}{s},\frac{(\beta_1^2+4\psi_i)^{1/2}}{s}p_{\textrm{inj}}^s\right]\right\}.
\nonumber \\
\label{MomSolDeltabiggerpi}
\end{eqnarray}

Because of the given structure of the solution
Eq.(\ref{MomSolGeneral}), it is convenient to introduce the momentum
value $p_{\star}$ via
\begin{equation}
p_{\star}=\left(\frac{s}{\sqrt{\beta_1^2+4\psi_i}}\right)^{1/s}.
\label{pstar}
\end{equation}
This parameter indicates the momentum value above which the cosmic
ray spectrum cuts off exponentially due to adiabatic losses
($\beta_1$) and escape losses ($\psi_i$). Now we can give some
asymptotic behaviour for the solution:
\\
In the case of low values of the momentum ($p \ll p_{\star}$), we
find, according to \citet{Abramowitz},
\begin{equation}
R_i(p \ll p_{\star})\simeq p^{+s-3},
\label{pkleinerpstern}
\end{equation}
which is a flat power-law spectrum at very small momenta. For large
arguments ($p \gg p_{\star}$) of the Kummer function in
Eq.(\ref{MomSolDeltabiggerpi}), we get
\begin{equation}
R_i(p \gg p_{\star}) \propto p^{\frac{a-3}{2}+\frac{a+3}{2}\frac{\beta_1}{\sqrt{\beta_1^2+4\psi_i}}-\frac{\xi_1}
{\sqrt{\beta_1^2+4\psi_i}}} \exp\left[ -\frac{\beta_1+(\beta_1^2+4\psi_i)^{1/2}}{2s}p^s\right].
\label{pgroesserpstern}
\end{equation}
In this situation, that the exponential cutoff is not negligible. In
the special case of dominating adiabatic losses, we take $\beta_1^2
\gg 4\psi_i$ to get
\begin{equation}
R_i\left( p \gg (s/\beta_1)^{1/s}\right)=p^{+a-\xi_1/\beta_1} \exp\left[ -\frac{\beta_1}{s}p^s\right].
\label{pgroesserpsternadiabatic}
\end{equation}
Also in this case the exponential cutoff dominates the spectral
behavior. To summarize these results we expect a flat power-law
spectrum at very low momenta, whereas the exponential cutoff
dominates at very high momenta.
 \begin{figure}
   \centering
   \includegraphics[width=10cm]{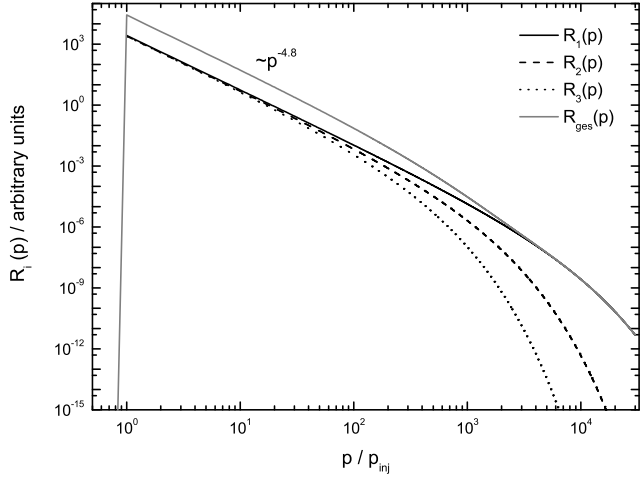}
      \caption{Cosmic ray spectrum (grey line) for a delta shape injection at $p=p_{\textrm{inj}}$ obtained by
 superposing
the contributions from individual modes ($i=1,2,3$). The first mode
$R_1(p)$ modelling the longest escape losstime is responsible for
the high energy cutoff whereas the slope of the power law is
determined by the most effective energy loss process (adiabatic
loss).
              }
         \label{fig:MomSol}
   \end{figure}

As an illustrative example a spectrum is modelled such that the
result is a power law-spectrum with an exponent like the one
observed from high energetic cosmic rays on Earth. This is
demonstrated in Fig.(\ref{fig:MomSol}). Here particles are injected
at a momentum value of $p=p_{\textrm{inj}}$ with a delta-shape
injection function. Furthermore we assume an isotropic Kolmogorov
turbulence model, i.e., the power law index of the turbulence is
$q=5/3$. We found a steeper power-law spectrum as the predicted one
from Eq.(\ref{pkleinerpstern}) over about two decades in momentum.
In this regime the exponential cutoff already plays a nonvanishing
role. As the result of the small momentum dependence of the
exponential term in Eq.(\ref{MomSolDeltasmallerpi}) ($s\equiv 1/3$
for Kolmogorov-like turbulence), a power-law spectrum over just two
decades in momentum is obtained. Performing some delta-shape
injections at increasing momenta with adequate normalisation values
a power-law spectrum over a wide range in momentum is obviously
possible.

The parameter that affects the final power-law index of the cosmic
ray spectrum is the parameter $a$ that describes the ratio of energy
gains from the Fermi I process and the Fermi II momentum diffusion
process. To match the $-4.8$ spectrum from observations, the value
has been chosen as $a=250$. The adiabatic losses are assumed to be
as strong as the Fermi II energy gains. In this context we need to
remember that the solution function $R_i(p)$ has to be multiplied by
$4\pi p^2$ to obtain the number of particles at the momentum $p$:
\begin{equation}
N_i(p)=4\pi p^2 R_i(p). \label{MomentumNumberDensity}
\end{equation}
The spectrum shown in Fig.(\ref{fig:MomSol}) comprises the first 10
eigenfunctions, and the cumulative solution (grey line) is given by
$R_{tot}(p)=R_1(p)+ R_2(p)+...+R_{10}(p)$. It can be seen that the
spectrum is almost completely dominated by the first eigenfunction.
Especially the cutoff is dominated by this first Eigenfunction. That
result has to be compared with, e.g., Fig. 2 of \citet{LS88}, where
the situation in spiral galaxies is described. Both, elliptical and
spiral galaxies have in common that the the cutoff is described by
the first Eigenfunction. The difference between these two situations
is that, at low energies, the higher Eigenfunctions describe the
spectrum in spiral galaxies. The resulting power law is made by the
sum of the single Eigenfunctions. In elliptical galaxies the
situation is different, because the spectral shape is by oneself
dominated by the first Eigenfunction.


The reason for this different behavior compared to spiral galaxies
is based on the loss time scales. While in the latter the escape
loss processes dominate because of the small galactic height
compared to the radial size, this is different in elliptical
galaxies, where the smallest `edge' of the confinement region is
given by the semi-minor axis being larger than the galactic height
in spirals. Therefore the dominating loss process in elliptical
galaxies is adiabatic loss.
\section{Connection to observations}
\label{ConnectionToObservations}
The results of the solutions of the transport equations have been explained in the previous section. For astrophysical scenarios it is important
to compare these with observations. Here we concentrate on the gamma-ray spectrum due to neutral pion decay and on the escaping cosmic ray spectrum from elliptical galaxies.
\subsection{Gamma-ray flux from pion decay}
\label{GammaFluxPion} The found solutions allow us to calculate the
gamma-ray flux from pion decay. The pions are mainly produced due to
proton-proton interactions, so we concentrate on the reaction
$p+p\rightarrow p+p+x\times \pi^0$. The neutral pion decays after
the mean lifetime $\tau=9\times 10^{-17}$s in two gamma-photons.
Other interaction channels can be treated analogously. Then the
derivation is straightforward, but tedious.
\\
We have to convolve the high-energy proton spectrum,
Eq.(\ref{GenSol}) multiplied with Eq.(\ref{MomSolDeltabiggerpi}),
with the pion power of a single proton. The latter one is
approximately given by \citep[cf.][]{MannSchlicki, Schlick1}:
\begin{equation}
P_{\pi}(\gamma_{\pi},\gamma_{p},\xi,\eta,\phi)=
c\gamma_{\pi}T(\xi,\eta,\phi)\Xi\sigma_{pp}^{\pi^0}(\gamma_p)\times \delta(\gamma_{\pi}-\gamma_{p}^{3/4})H\left[\gamma_p-\gamma_{\textrm{thr}}\right].
\label{PionLeistung}
\end{equation}
In this equation, $\Xi\sigma_{pp}^{\pi^0}(\gamma_p)$ gives the
multiplicity $\Xi$ for neutral pion production in p-p interactions
and $\sigma_{pp}^{\pi^0}(\gamma_p)$ the total cross-section. Within
the high energy limit, this factor is given by
\citep[cf.][]{Schlick1}:
\begin{equation}
\Xi\sigma_{pp}^{\pi^0}(\gamma_p)=\sigma_{0,pp}^{\pi^0}(\gamma_p-1)^{0.53}.
\label{Multiplizitaet}
\end{equation}
The factor $\gamma_{\textrm{thr}}$, together with the Heaviside
function, gives the threshold value for the proton energy to produce
neutral pions ($\gamma_{\textrm{thr}}m_pc^2=1.22$GeV). Therefore we
get, for the pion source function,
\begin{eqnarray}
Q_{\pi^0}(\gamma_{\pi},\xi,\eta,\phi)&=&\frac{1}{\gamma_{\pi}m_{\pi}c^2}\int_1^{\infty}\,d\gamma_p
T(\xi,\eta,\phi)C_{\textrm{inj}}\gamma_p^{a+2}
\exp\left[ -\frac{\beta_1+(\beta_1^2+4\psi_i)^{1/2}}{2s}(m_{p}c)^s\gamma_{p}^s\right]\times
\nonumber \\
&&\times U\left[\frac{a+3}{2s}+\frac{\xi_1-[(3+a)/2]\beta_1}{s(\beta_1^2+4\psi_i)
^{1/2}},\frac{a+3}{s},\frac{(\beta_1^2+4\psi_i)^{1/2}}{s}(m_{p}c)^s\gamma_{p}^s\right]\times
\nonumber \\
&&\times
c\gamma_{\pi}T(\xi,\eta,\phi,u)\sigma_{0,pp}^{\pi^0}(\gamma_p-1)^{0.53}\times
\delta(\gamma_{\pi}-\gamma_{p}^{3/4})H\left[\gamma_p-\gamma_{\textrm{thr}}\right].
\label{PionSource}
\end{eqnarray}
The power-law factor $p^{a+2}$ is due to Eq.(\ref{MomentumNumberDensity}) and we used
\begin{eqnarray}
C_{\textrm{inj}}&=&4\pi(m_pc)^{a+2}
\times(a_2p_1^s)^{-1}\frac{\Gamma\left[\frac{a+3}{2s}+\frac{\xi_1-[(3+a)/2]\beta_1}{s(\beta_1^2+\psi_i)
^{1/2}} \right]}{\Gamma\left[\frac{a+3}{s} \right]} \times \left[\frac{\beta_1^2+4\psi_i}{s^2}\right]
^{\frac{a+3}{2s}} \times
\nonumber \\
&&\times p_{\textrm{inj}}^{s+1}\exp\left[\frac{\beta_1+(\beta_1^2+4\psi_i)^{1/2}}{2s}
p_{\textrm{inj}}^s\right]\times
\nonumber\\
&&\times
M\left[\frac{a+3}{2s}+\frac{\xi_1-[(3+a)/2]\beta_1}{s(\beta_1^2+4\psi_i)
^{1/2}},\frac{a+3}{s},\frac{(\beta_1^2+4\psi_i)^{1/2}}{s}p_{\textrm{inj}}^s\right].
\label{Ckonstante}
\end{eqnarray}
Performing the integration we get
\begin{eqnarray}
Q_{\pi^0}(\gamma_{\pi},\xi,\eta,\phi)&=&\frac{4}{3}\frac{T^2(\xi,\eta,\phi)\sigma_{0,pp}^{\pi^0}}{m_{\pi c}}\times
\nonumber \\
&&\times C_{\textrm{inj}}\gamma_{\pi}^{4a/3+3}(\gamma_{\pi}^{4/3}-1)^{0.53}\exp\left[ -\frac{\beta_1+(\beta_1^2+4\psi_i)^{1/2}}{2s}(m_{p}c)^s\gamma_{p}^{4s/3}\right]\times
\nonumber \\
&&\times U\left[\frac{a+3}{2s}+\frac{\xi_1-[(3+a)/2]\beta_1}{s(\beta_1^2+4\psi_i)
^{1/2}},\frac{a+3}{s},\frac{(\beta_1^2+4\psi_i)^{1/2}}{s}(m_{p}c)^s\gamma_{p}^{4s/3}\right]\times
\nonumber \\
&&\times H\left[\gamma_{\pi}-\gamma_{\textrm{thr}}^{3/4}\right].
\label{PionSDeltaInj}
\end{eqnarray}
It is obvious that the pion source function is proportional to the
squared spatial distribution of the high-energy protons since pion
production is a two-body process. Therefore the brightest luminosity
is strongly correlated with the highest proton density in spatial
coordinates because the mean lifetime of the neutral pion is
extremely short so that the photons are produced very close to the
p-p interaction region.
\\
With this result, the differential gamma-ray source function is
given by  \citep[cf.][]{Schlick1}
\begin{equation}
Q_{\gamma}(E_{\gamma},\xi,\eta,\phi)=2\int_{\gamma_{\textrm{min}}}^{\infty}\,d\gamma_{\pi}\frac{Q_{\pi^0}
(\gamma_{\pi},\xi,\eta,\phi)}
{\sqrt{\gamma_{\pi}-1}},
\label{GammaIntDeltaInj}
\end{equation}
where
$\gamma_{\textrm{min}}=\frac{E_{\gamma}}{m_{\pi}c^2}+\frac{m_{\pi}c^2}{4E_{\gamma}}$
denotes the minimum energy of the gamma-ray photons after the decay
of the neutral pion. We can state that the differential gamma-ray
source function in the high-energy regime ($E_{\gamma} \gg
m_{\pi^0}c^2$), where the spectrum shows a power-law dependence, is
approximately given by
\begin{eqnarray}
Q_{\gamma,\pi^0}(E_{\gamma},\xi,\eta,\phi)&\propto&
\gamma_{\pi}^{4a/3+3}(\gamma_{\pi}^{4/3}-1)^{0.53}\exp\left[ -\frac{\beta_1+(\beta_1^2+4\psi_i)^{1/2}}{2s}(m_{p}c)^s\gamma_{p}^{4s/3}\right]\times
\nonumber \\
&&\times U\left[\frac{a+3}{2s}+\frac{\xi_1-[(3+a)/2]\beta_1}{s(\beta_1^2+4\psi_i)
^{1/2}},\frac{a+3}{s},\frac{(\beta_1^2+4\psi_i)^{1/2}}{s}(m_{p}c)^s\gamma_{p}^{4s/3}\right],
\label{GammaDifferentiellDeltaInj}
\end{eqnarray}
but the integral Eq.(\ref{GammaIntDeltaInj}) cannot be performed
analytically.
\\
Consequently, as discussed before, we assume a power-law spectrum
over a wide range in momentum until a maximum value of
$\gamma_{\textrm{p,\,max}}m_pc^2$:
\begin{equation}
N(\gamma_p)=N_0\gamma^{-z}H\left[\gamma_{\textrm{p,\,max}}-\gamma_p
\right]. \label{PowerLawGammaMomentum}
\end{equation}
After performing the convolution (cf. Eq.(\ref{PionSource})) the
pion source function reads
\begin{equation}
Q_{\pi^0}(\gamma_{\pi},\xi,\eta,\phi)=\frac{4}{3}\frac{T^2(\xi,\eta,\phi)\sigma_{0,pp}^{\pi^0}}
{m_{\pi c}}
\gamma_{\pi}^{-(4z-1)/3}(\gamma_{\pi}^{4/3}-1)^{0.53}H\left[\gamma_{\pi}-\gamma_{\textrm{thr}}^{3/4}\right]
H\left[\gamma_{\textrm{p,\,max}}^{3/4}-\gamma_{\pi}\right].
\label{PionSourcePowerLaw}
\end{equation}
From this calculation we get due to Eq.(\ref{GammaIntDeltaInj}) in
the high-energy limit ($\sqrt{\gamma_{\pi}^2-1}\simeq \gamma_{\pi}$
and $(\gamma_{\pi}^{4/3}-1)^{0.53}\simeq \gamma_{\pi}^{4/3\times
0.53}$) for gamma-ray energies above $130$ GeV
\begin{equation}
Q_{\gamma}(E_{\gamma},\xi,\eta,\phi)\simeq \frac{8}{3}\frac{N_{0}T^2(\xi,\eta,\phi)\sigma_{0,pp}^{\pi^0}}{m_{\pi}c}
\left[ \left(1.04-\frac{4}{3}z\right)^{-1}\left(\gamma_{\textrm{p,\,max}}^{-z+0.78}-\gamma_{\textrm{min}}^
{-4/3z+1.04}\right)\right]
\label{GammaRayPionPowerLaw}
\end{equation}
as the gamma-ray source function. Thus the differential gamma-ray
source function at high energies ($E_{\gamma} \gg m_{\pi^0}c^2$) is
given by the relation
\begin{equation}
Q_{\gamma,\pi^0}(E_{\gamma},\xi,\eta,\phi)\propto E_{\gamma}^{-s_{\pi}}=E_{\gamma}^{-4/3z+1.04}
\label{GammaDifferentiellPowerLawIndex}
\end{equation}
after Eq.(\ref{PionSourcePowerLaw}). Taking an injection power-law
index $z=2.8$, we find that
$Q_{\gamma,\pi^0}(E_{\gamma},\xi,\eta,\phi)\propto
E_{\gamma}^{-2.69}$, which is slightly harder than the injected
proton spectrum. Therefore an observed gamma-ray flux from an
elliptical galaxy gives constraints to the high-energy cosmic ray
spectrum within that object since neutral pions are produced in the
whole confinement volume. Therefore we expect an elliptical galaxy
be an extended source of gamma rays.
\subsection{Escape-spectrum of cosmic rays}
We next derive the spectrum leaking out of elliptical galaxies.
Therefore the mean free path $\lambda_{\textrm{mfp}}(p)$ for single
scattering events between charged particles and plasma waves is
needed. According to \citet{Schlick1}, this characteristic length is
given by
\begin{equation}
\lambda_{\textrm{mfp}}(p)\equiv\frac{3K(\textbf{r},p)}{v}=\lambda_{\textrm{0,\,mfp}}p^{2-q},
\label{lambdameanfreepath}
\end{equation}
where $K(\textbf{r},p)$ and $v$ denote the spatial diffusion coefficient and the particle velocity
respectively. Again the parameter $q$ gives the momentum dependence of the turbulence spectrum.
Here we use $v=c$ for highly relativistic cosmic rays.
 \begin{figure}
   \centering
   \includegraphics[width=10cm]{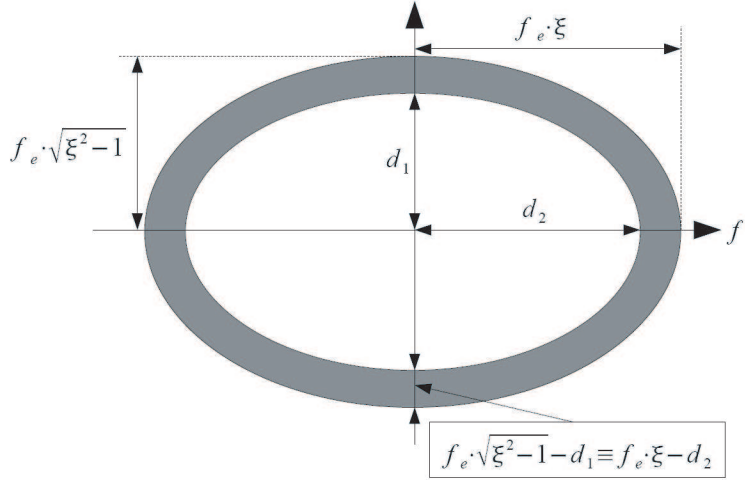}
      \caption{Schematic view of the geometry for calculating the escape-spectrum
      from elliptical galaxies. Particles within the gray-shaded area leave the galaxy by chance,
       if their mean free path $\lambda_{\textrm{mfp}}(p)$ is longer than the distance
        to the edge of the galaxy.
              }
         \label{fig:EscapeSpectrum}
   \end{figure}
Figure(\ref{fig:EscapeSpectrum}) gives the chosen geometry for
calculating of the escape spectrum. We assume that cosmic rays will
escape out of the galaxy, if their mean free path is larger than the
distance to the edge of the galaxy. This border value is given by
(cf. Ch.(\ref{MomentumSolution}))
\begin{equation}
r_g(\xi,\eta,\phi)=f_e\sqrt{(\xi\eta)^2+(\xi^2-1)(1-\eta^2)}.
\label{DefRadius}
\end{equation}
All particles in the gray-shaded area will escape the galaxy, if
$r_g(\xi,\eta,\phi)-d_1\leq\lambda_{\textrm{mfp}}(p)$, where we
neglect a geometrical factor of order unity. The width of this
spherical shell is constant in every direction so that
$f_e\sqrt{\xi^2-1}-d_1\equiv f_e\xi-d_2$, as indicated in
Fig.(\ref{fig:EscapeSpectrum}). The rate of the escaping high
energetic particles is approximatively given by the inverse escape
time
\begin{equation}
\frac{K(\textbf{r},p)}{r^2(\xi,\eta,\phi)}=\frac{c\lambda_{\textrm{mfp}}(p)}{3f^2\left[(\xi\eta)^2+(\xi^2-1)(1-\eta^2)\right]},
\label{Entweichrate}
\end{equation}
where we neglect the geometrical factor for the direction of the
trajectory of cosmic ray particles. Therefore the total number of
particles escaping out from the gray-shaded area into the
intergalactic medium per unit time and per unit momentum is given by
\begin{eqnarray}
N(p)&\propto&4\pi p^2\int_0^{2\pi}\int_{-1}^{1}\int_{d_1}^{r_g(\xi,\eta,\phi)}\sum_i\,\alpha_iT_i(\xi,\eta,\phi)R_i(p)\times
\nonumber \\
&&\times H\left[\lambda_{\textrm{mfp}}(p)-(r_g(\xi,\eta,\phi)-d_1)\right]
\frac{c\lambda_{\textrm{mfp}}(p)}{3f^2\left[(\xi\eta)^2+(\xi^2-1)(1-\eta^2)\right]}f^2df
d\eta d\phi.
\label{EscapeIntegral}
\end{eqnarray}
The upper boundary in the $f$-integral gives the radius of the
galaxy with respect to the variables $\xi$ and $\eta$. As an
illustration we choose a constant cosmic-ray distribution like
Eq.(\ref{TSolution}) with Eq.(\ref{diTildeXiL})(see Appendix B).
This one is independent of the variables $\eta$ and $\phi$. More
realistic particle distribution functions can be treated
analogously. Furthermore, we take a constant value of $\xi$ for a
dedicated elliptical galaxy. The Heaviside function $H\left[
\lambda_{\textrm{mfp}}(p)-(r_g(\xi,\eta,\phi)-d_1)\right]$ implies
that only high energetic particles with
$\lambda_{\textrm{mfp}}(p)\geq r_g(\xi,\eta,\phi)-d_1$ can leave the
galaxy.

Under these conditions we can convert Eq.(\ref{EscapeIntegral}) into
\begin{equation}
N\propto 4\pi
p^2\int_{0}^{2\pi}\int_{-1}^{1}\int_{r_g(\xi,\eta,\phi)-\lambda_{\textrm{mfp}}(p)}^{r_g(\xi,\eta,\phi)}
\sum_i\,\alpha_iT_i(\xi=const)R_i(p)\frac{c\lambda_{\textrm{mfp}}(p)}{3\left[(\xi\eta)^2+(\xi^2-1)(1-\eta^2)\right]}dfd\eta
d\phi. \label{EscapeIntegralVereinfacht}
\end{equation}
The general solution of this integral is given by
\begin{equation}
N\propto\frac{16\pi^2cp^2\lambda_{\textrm{mfp}}^2(p)\,\cot^{-1}(\sqrt{\xi^2-1})}{3\sqrt{\xi^2-1}}\sum_i\alpha_iT_i(\xi=const)R_i(p).
\label{IntegralGeloest}
\end{equation}
From this calculation we recognise that the escaping spectrum we
observe depends on the shape of the galaxy given by the choice of
the parameter $\xi$=constant. Inserting
Eq.(\ref{lambdameanfreepath}) into Eq.(\ref{IntegralGeloest}), we
get
\begin{equation}
N\propto \frac{16}{3}\pi^2c\lambda_{\textrm{0,\,mfp}}^2\sum_i\alpha_iT_i(\xi=const)R_i(p)p^{6-2q}.
\label{EscapeSpectrumFuehrend}
\end{equation}
Under the assumption that $R_{\textrm{ges}}(p)$ is proportional to
$p^{-4.8}$, we find for a Kolmogorov turbulence model ($q=5/3$) a
power-law index of $-2.13$ for the escaping spectrum. This spectral
behaviour is harder than the intrinsic spectrum, which has a
spectral index of $-2.8$ (cf. Ch.(\ref{MomentumSolution})).

Such an effect can be easily understood by the high energetic
particles leaving the galactical confinement region more frequently
than low energetic ones. As a consequence there are two
possibilities explaining the observed high energy cosmic ray
spectrum above the `knee' with spectral index of $-3.1$ assuming
that elliptical galaxies provide a significant amount to the overall
high-energy cosmic ray flux. First, the intrinsic spectrum in
elliptical galaxies may be steeper then that one we have chosen
here. In this context the measurement of the gamma-ray flux reaching
the Earth from such an object would give us interesting constraints.
Second, if the intrinsic spectrum in elliptical galaxies has nearly
the same dependence as the one we measure here on Earth
($N(p)\propto p^{-2.7}$), the transport of cosmic rays after escape
from elliptical galaxies depends on energy. In these two scenarios
it is possible to explain the high-energy component with our model.

Note that, due to the previously given arguments, the highest
energetic particles cannot be confined within the galactical volume.
In our examination the maximum energy of confined charged particles
within elliptical galaxies is given by the relation
\begin{equation}
f_e\xi=\lambda_{\textrm{mfp}}(p)=\lambda_{\textrm{0,\,mfp}}p^{2-q}
\label{MaxEnergy}.
\end{equation}
For reasonable values of giant elliptical galaxies
($\lambda_{\textrm{mfp}}(p=10\, \textrm{GeV}\,
\textrm{c}^{-1})=10^{19}$cm, $f_e\xi=5\times 10^{22}$ cm,
\citep[cf.][]{Schlick1}) we get from Eq.(\ref{MaxEnergy}) a value of
about $10^{21}$ eV for cosmic ray protons. Above this energy value,
it is generally impossible that elliptical galaxies confine cosmic
rays within their volume.


\section{Conclusion}

We showed that an analytical treatment of cosmic ray transport in elliptical
galaxies based on the diffusion approximation is possible in general. The
formal solution we found, combined with appropriate boundary conditions and
source functions, can be used to study transport processes in elliptical
galaxies. This model is valid for the complete physical parameter space as long
as the diffusion approximation holds.

The first test case where we applied our model to is a jet-like
injection shape like in M87. As we can separate our problem into
spatial and momentum problems, this test case probes the spatial
problem. For this model we found that for very long timescales
cosmic rays are distributed throughout the galaxy almost
isotropically.

We also provided a test case for the momentum problem. Under the
assumption of a resulting power-law spectrum matching the observed
power law-spectrum on Earth, we could identify the governing
eigenfunctions and therefore the governing processes for cosmic ray
acceleration and momentum diffusion in  elliptical galaxies. It
turned out that in elliptical galaxies adiabatic losses are
responsible for the high energy cutoff, whereas the slope of the
spectrum is given by the ratio of the strength of Fermi I and Fermi
II processes. As a result of the basic possibility of explaining
this power law-with physical parameters that might be found in M87,
our calculation gives rise to the theory of M87 as a source of
ultra-high energy cosmic rays. In this context the gamma-ray flux
from pion decay and the escaping cosmic ray spectrum are essential
for a more adequate view of cosmic ray astrophysics. We find that
the gamma-ray spectrum with a power law index of $-2.69$ is a bit
harder than the intrinsic cosmic ray spectrum (power law index
$-2.8$). In the context of understanding the sources of high-energy
cosmic rays, the flux of charged particles escaping from the
confinement region of elliptical galaxies is also very interesting.
For the same intrinsic spectrum as above, we find $-2.13$ as the
spectral index for escaping particles. Again the spectrum is harder
than the value within elliptical galaxies.
\\
Our model may also serve as a testbed for some more advanced
questions:
\begin{itemize}
\item Are elliptical galaxies, and especially AGN within these objects, reasonable sources of the ultrahigh energy cosmic rays?
\item What is the total flux leaving a single elliptical galaxy with respect to a measured source of cosmic rays in elliptical galaxies?
\item What is the total cosmic ray flux we can expect to measure on Earth in the special case of
the nearby active elliptical galaxy M87 or Centaurus A?
\end{itemize}
Clearly, to answer these questions, a detailed quantitative analysis
of the properties of cosmic ray transport within elliptical galaxies
is required. Therefore we need to determine from observations the
exact physical parameters of the interstellar medium and of the
emission processes within the source. With our analytical model at
hand, the task of constraining the possible parameter space gets
easier. Especially by assuming that UHECR are coming mostly from
sources like M87, we may be able to derive the total cosmic ray
content of this type of elliptical galaxies.

Nevertheless this work is considered as a starting point for more
sophisticated (numerical) models. Another interesting point is the
expansion of this model to cosmic ray electrons. While the spatial
description may be derived analogously, we need to add the physical
processes involved in the transport of electrons in momentum space.
Synchrotron losses and the inverse Compton effect will then play a
major role \citep[see, e.g.,][]{Casadei}. The output of this model
can be easily tested with the radio data of elliptical galaxies
since the electrons provide the main contribution to this radiation.
\begin{acknowledgements}
      TH acknowledges support by Graduiertenkolleg 1147 and FS acknowledges support
      by the
      \emph{Deut\-sche For\-schungs\-ge\-mein\-schaft, DFG\/} project
      number Sp~1124/1--1. We would like to thank R. Schlickeiser for his useful comments.
\end{acknowledgements}


\section*{Appendix A: General solution of Eq.(\ref{DGL})}
The general solution of Eq.(\ref{DGL}) can be obtained by consecutive separation of variables.
First we use
\begin{equation}
T(\xi,\eta,\phi,u)=X(\xi,\eta,\phi)U(u), \label{SepariertU}
\end{equation}
leading to the following differential equations with $-k^2$ as the
separation constant:
\begin{equation}
\frac{\partial U}{\partial u}=-k^2 U \label{DGLU}
\end{equation}
\begin{eqnarray}
&&\frac{\partial}{\partial \xi}\left[(\xi^2-1)\frac{\partial
X}{\partial \xi}\right]+\frac{\partial}{\partial
\eta}\left[(1-\eta^2)\frac{\partial X}{\partial \eta}\right]+
\nonumber \\
&&+\frac{\xi^2-\eta^2}{(\xi^2-1)(1-\eta^2)}\frac{\partial^2
X}{\partial \phi^2}+c^2(\xi^2-\eta^2)X=0. \label{DGLX}
\end{eqnarray}
The factor $c$ is given by
\begin{equation}
c=\frac{fk}{\sqrt{K_0}}. \label{cFokenabstand}
\end{equation}
Equation (\ref{DGLU}) is easily solved by
\begin{equation}
U(u)=\exp(-k^2 u), \label{LoesungU}
\end{equation}
where we put the integration constant to equality without loss of generality. This result serves an a
consistency check to the eigenfunction expansion Eq.(\ref{EigExp}). The separation constant $k$ will be
calculated by use of spatial boundary conditions.
\newline
Eq.(\ref{DGLX}) can be solved through another separation of variables. With
\begin{equation}
X(\xi,\eta,\phi)=A(\xi,\eta)B(\phi), \label{SepariertPhi}
\end{equation}
one gets
\begin{eqnarray}
&&-\frac{1}{B}\frac{\partial^2 B}{\partial \phi^2}=
\frac{(\xi^2-1)(1-\eta^2)}{(\xi^2-\eta^2) \times A}
\left\{\frac{\partial}{\partial \xi}\left[(\xi^2-1)\frac{\partial
A}{\partial \xi}\right]+ \right.
\nonumber \\
&&\left.+\frac{\partial}{\partial
\eta}\left[(1-\eta^2)\frac{\partial A}{\partial \eta}\right]\right\}
+(\xi^2-1)(1-\eta^2)c^2.  \label{DGLsepariert2}
\end{eqnarray}
The separation constant is set to be $+m^2$. Therefore we get
\begin{equation}
\frac{\partial^2 B}{\partial \phi^2}+m^2 B=0, \label{DGLB}
\end{equation}
and
\begin{eqnarray}
&&\frac{(\xi^2-1)(1-\eta^2)}{(\xi^2-\eta^2) \times A}
\left\{\frac{\partial}{\partial \xi}\left[(\xi^2-1)\frac{\partial
A}{\partial \xi}\right]+\frac{\partial}{\partial
\eta}\left[(1-\eta^2)\frac{\partial A}{\partial
\eta}\right]\right\}+
\nonumber \\
&&+\left\{(\xi^2-1)(1-\eta^2)c^2-m^2\right\}A = 0.  \label{DGLA}
\end{eqnarray}
The solution of Eq.(\ref{DGLB}) is given by
\begin{equation}
B(\phi)=\sin(m\phi)+ \cos(m\phi),
\label{LoesungPhi}
\end{equation}
where we again used integrating constants equal to $1$. The last
separation
\begin{equation}
A(\xi,\eta)=R(\xi)S(\eta) \label{LastSeparation}
\end{equation}
solves Eq.(\ref{DGLA}), which can be written by
\begin{eqnarray}
&&\frac{\partial}{\partial \xi}\left[(\xi^2-1)\frac{\partial
A}{\partial \xi}\right]+\frac{\partial}{\partial
\eta}\left[(1-\eta^2)\frac{\partial A}{\partial \eta}\right]+
\nonumber \\
&&+\left[(\xi^2-\eta^2)c^2-m^2
\left(\frac{1}{\xi^2-1}+\frac{1}{1-\eta^2}\right)\right]A=0.
\label{Partialfrac}
\end{eqnarray}
With the separation constant $+\beta_{mn}$ we get
\begin{equation}
\frac{\partial}{\partial \xi}\left[(\xi^2-1)\frac{\partial
R}{\partial
\xi}\right]-\left(\beta_{mn}-\xi^2c^2+\frac{m^2}{\xi^2-1}\right)R=0,
\label{DGLR}
\end{equation}
and
\begin{equation}
\frac{\partial}{\partial \eta}\left[(1-\eta^2)\frac{\partial
S}{\partial
\eta}\right]+\left(\beta_{mn}-\eta^2c^2-\frac{m^2}{1-\eta^2}\right)S=0.
\label{DGLS}
\end{equation}
It is notable that the $\xi$ and $\eta$ solution functions satisfy
similar differential equations. From \citet{Abramowitz} we find
\begin{eqnarray}
R_{mn}^{(p)}(c,\xi)&=& \left[\sum_{r=0,1}^{\infty ^{\prime}}
\frac{(2m+r)!}{r!}d_r^{mn}\right]^{-1}\times \left(\frac{\xi^2-1}{\xi^2}\right)^{m/2} \times
\nonumber \\
& &\times \sum_{r=0,1}^{\infty ^{\prime}} i^{r+m-n}
\frac{(2m+r)!}{r!}d_r^{mn} Z_{m+r}^{(p)}(c\xi) \label{LoesungXi}
\end{eqnarray}
as the solution for the $\xi$ dependence of Eq.(\ref{DGL}). It is
$Z_{n}^{(p)}(c\xi)=\sqrt{\frac{\pi}{2c\xi}}J_{n+\frac{1}{2}}(c\xi)$
for the first kind ($p=1$) and
$Z_{n}^{(p)}(c\xi)=\sqrt{\frac{\pi}{2c\xi}}Y_{n+\frac{1}{2}}(c\xi)$
for the second kind ($p=2$). $J$ and $Y$ are Bessel functions of the
first and second kind respectively. For our model we can neglect the
solution of second kind because of physical motivated spatial
boundary conditions (see Appendix B). It is clear that $n=m+r$. So
$i^{r+m-n}$ in Eq.(\ref{LoesungXi}) is equal to $1$. The mark
$^{\prime}$ connected with the sum means a summation either over
even or over odd values of $r$. The latter ones as well as the
weighting factors $d_r^{mn}$ have also been adjusted to spatial
boundary conditions.
\newline
Eq.(\ref{DGLS}) has the formal mathematical solution given by
\begin{equation}
S_{mn}^{(1)}(c,\eta)=\sum_{r=0,1}^{\infty
^{\prime}}\hat{d}_r^{mn}(c)P_{m+r}^m(\eta), \label{Loesung1Eta}
\end{equation}
and
\begin{equation}
S_{mn}^{(2)}(c,\eta)=\sum_{r=0,1}^{\infty
^{\prime}}\hat{d}_r^{mn}(c)Q_{m+r}^m(\eta). \label{Loesung2Eta}
\end{equation}
Here $P_{m+r}^m(\eta)$ and $Q_{m+r}^m(\eta)$ denote associated
Legendre polynomials of first and second kind respectively and
$\hat{d}_r^{mn}(c)$ are weighting factors. Again $^{\prime}$ means a
summation over even or odd values of $r$. We can neglect the
solution function of second kind with respect to spatial boundary
conditions (see the consistency checks).
\newline
Under these conditions we can write the formal mathematical solution
as
\begin{eqnarray}
T(\xi,\eta,\phi, u)&=&\sum_{m,n} R_{mn}^{(1)}(c,\xi)\times S_{mn}^{(1)}(c,\eta)\times
\cos(m\phi)\times \exp(-k^2u)=
\nonumber \\
&=&\sum_{m,n} \left\{\left[\sum_{r=0,1}^{\infty ^{\prime}}
\frac{(2m+r)!}{r!}d_r^{mn}\right]^{-1}\times \left(\frac{\xi^2-1}{\xi^2}\right)^{m/2} \times\right.
\nonumber \\
& &\left.\times \sum_{r=0,1}^{\infty ^{\prime}}
\frac{(2m+r)!}{r!}d_r^{mn} \sqrt{\frac{\pi}{2c\xi}}J_{n+\frac{1}{2}}(c\xi) \times\right.
\nonumber \\
& &\left. \times\hat{d}_r^{mn}(c)P_{m+r}^{m}(\eta) \times
\cos(m\phi)\times \exp(-k^2u) \right\}. \label{GenSolAppend}
\end{eqnarray}

\section*{Appendix B: Effect of spatial boundary conditions to the general solution Eq.(\ref{GenSol})}

As an illustrative example we calculate the solution depending on
the following boundary conditions. First we define one for the
variable $\xi$. Remembering, that $f\xi$ is equal to the semi-major
axis $a$ we assume `free escape' boundary conditions in the form
\begin{equation}
T(\xi=\xi_{c},\eta,\phi,u)=0. \label{BoundXi}
\end{equation}
Hence the particle distribution function $T(\xi,\eta,\phi,u)$ is set
to be zero at the maximum size $f\xi_c$ of the confinement region of
cosmic ray particles being a convenient way to model leakage out of
the galaxy. This region is assumed to be larger than the (optical)
size of the galaxy ($f\xi_g$). Furthermore we assume periodical
boundary conditions for $\eta$ and $\phi$:
\begin{equation}
T(\xi,\eta = \-1,\phi,u)=T(\xi,\eta = +1,\phi,u) \label{BoundEta}
\end{equation}
\begin{equation}
T(\xi,\eta,\phi=2\pi,u)=T(\xi,\eta,\phi=0,u) \label{BoundPhi}
\end{equation}
Finally we take
\begin{equation}
T(\xi,\eta,\phi,u=0)=q_0
\Theta(\xi_{g}-\xi)\label{BoundU}
\end{equation}
for a given distance of the foci ($2f$) as a simple model to explain
a constant source function over the whole size of the galaxy. Here
we neglect the $\phi$- dependence (taking a $\phi$- dependence into
account is straightforward). As a consequence we set $m=0$ in
Eq.(\ref{GenSol}) leading to $n=r$ (cf. Appendix A).

In order to match Eq.(\ref{BoundEta}) only even solution functions
are practical. This is done by  $P_{r}^0(\eta)$. Therefore we
neglect $Q_{r}^0(\eta)$ having singularities at the points $\eta=1$
and $\eta=-1$. From Eq.(\ref{BoundXi}) we recognise that
$J_{r+\frac{1}{2}}(c\xi)=0$ if $\xi=\xi_{c}$. Let $y_{ri}$ be the
zeros of $J_{r+\frac{1}{2}}(\xi)$, then we find
\begin{equation}
c=\frac{y_{ri}}{\xi_{c}} \hspace{0.2cm},\hspace{0.2cm}
k_{ri}=\frac{y_{ri}\sqrt{K_0}}{\xi_{c}f}. \label{Nullstelleyi}
\end{equation}
As a result we write
\begin{equation}
k^2 \equiv k_{ri}^2=\frac{K_0 y_{ri}^2}{\xi_c^2 f^2}. \label{kri}
\end{equation}
This means that we have to sum over all zeros $y_{ri}$ aditionally.
The last boundary condition Eq.(\ref{BoundU}) is useful to determine the weighting factors
$d_{ri}$ and $\hat{d}_{ri}$ respectively.
From
\begin{equation}
\sum_{i=1}^{\infty} \left\{ \frac{\sum_{r=0,1}^{\infty^{\prime}}d_{ri}\times \sqrt{\frac{\pi
\xi_{c}}{2y_{ri}\xi}}J_{r+\frac{1}{2}}\left(y_{ri}\frac{\xi}{\xi_{c}}\right)}
{\sum_{r=0,1}^{\infty^{\prime}}d_{ri}}
\times \sum_{r=0,1}^{\infty^{\prime}}\hat{d_{ri}}P_r^0(\eta)\right\}
=q_0 \Theta(\xi_{g}-\xi)
\label{driEq1}
\end{equation}
We see that we can cancel the factors $d_{ri}$.

Adjusting Eq.(\ref{GenSol}) to the boundary condition
Eq.(\ref{BoundEta}) it becomes clear that only even values of $r$
match the required periodicity. Using the orthonormality relation
for Legendre polynomials
\begin{equation}
\int_{-1}^1P_r^0(\eta)P_{r^{\prime}}^0(\eta)d\eta=\frac{2}{2r+1}\delta_{rr^{^{\prime}}},
\label{OrthonormalityLegendre}
\end{equation}
the RHS of Eq.(\ref{driEq1})  evolves to be
\begin{equation}
q_0 \Theta(\xi_{g}-\xi) \int_{-1}^1 P_{r^{^{\prime}}}^0(\eta) d\eta.
\label{rhsdri}
\end{equation}
Performing the integral it is obvious that only  $r^{^{\prime}}=0$ gives a non-vanishing value
for $\int_{-1}^1 P_{0}^0(\eta) d\eta=2$. Consequently we set $r=r^{^{\prime}}=0$.
As a result all summations over $r$ disappear so there is only
\begin{equation}
\sum_{i=1}^{\infty}\hat{d_{i}} \sqrt{\frac{\pi
\xi_{c}}{2y_i\xi}}J_{\frac{1}{2}}\left(y_i\frac{\xi}{\xi_{c}}\right)
= q_0 \Theta(\xi_{g}-\xi)
\label{driEq2}
\end{equation}
left. Here we use the orthonormality relation for Bessel functions,
i.e.
\begin{equation}
\int_0^a J_{\nu}\left( \alpha_{\nu u}
\frac{\rho}{a}\right)J_{\nu}\left( \alpha_{\nu w}
\frac{\rho}{a}\right) \rho d\rho = \frac{a^2}{2}[J_{\nu
+1}^2(\alpha_{\nu u})]\delta_{uw}, \label{OrthonormalityBessel}
\end{equation}
to calculate the weighting factors. The Bessel function of second
kind diverges at $\xi=0$ and is therefore not appropriate for the
general mathematical solution (cf. Appendix A). Finally we find (we
substitute $\alpha_i \equiv \sqrt{\frac{\pi\xi_{c}}{2y_i}} \times
\hat{d_{i}}$)
\begin{equation}
E_i(\textbf{r}) \equiv T(\xi,u)=\sum_{i=1}^{\infty}\alpha_i
\frac{1}{\sqrt{\xi}}J_{\frac{1}{2}}\left(y_i\frac{\xi}{\xi_{c}}\right)
\exp \left[ -\frac{K_0 y_i^2}{\xi_c^2 f^2}u \right ]
\label{TSolution}
\end{equation}
with the weighting factors
\begin{equation}
\alpha_i=\frac{q_0
\int_1^{\xi_g}\xi^{3/2}J_{\frac{1}{2}}\left(y_i\frac{\xi}{\xi_{c}}\right)
d\xi} {\frac{\xi_c^2}{2}\left[ J_{\frac{3}{2}}(y_i)\right ]^2}.
\label{diTildeXiL}
\end{equation}
Again we see that the general solution of Eq.(\ref{DGL}) with respect to spatial boundary conditions
is indeed of the form of the eigenfunction expansion Eq.(\ref{EigExp}) with Eq.(\ref{ExpCoeffEigen}).

\bibliographystyle{aa} 
\bibliography{Paperbib} 

\begin{thebibliography}{30}
\expandafter\ifx\csname natexlab\endcsname\relax\def\natexlab#1{#1}\fi

\bibitem[{{Abramowitz} \& {Stegun}(1972)}]{Abramowitz}
{Abramowitz}, M. \& {Stegun}, I.~A. 1972, {Handbook of Mathematical Functions}
  (Handbook of Mathematical Functions, New York: Dover, 1972)

\bibitem[{{Arfken} \& {Weber}(2005)}]{Arfken}
{Arfken}, G.~B. \& {Weber}, H.~J. 2005, Materials and Manufacturing Processes

\bibitem[{{Bekki} \& {Shioya}(1998)}]{BekkiShioya}
{Bekki}, K. \& {Shioya}, Y. 1998, \apj, 497, 108

\bibitem[{{Biermann}(1995)}]{Bier}
{Biermann}, P.~L. 1995, Nuclear Physics B Proceedings Supplements, 43, 221

\bibitem[{{Blandford} \& {Ostriker}(1978)}]{Bland}
{Blandford}, R.~D. \& {Ostriker}, J.~P. 1978, \apjl, 221, L29

\bibitem[{{Casadei} \& {Bindi}(2004)}]{Casadei}
{Casadei}, D. \& {Bindi}, V. 2004, \apj, 612, 262

\bibitem[{{Cowsik}(1980)}]{Cow}
{Cowsik}, R. 1980, \apj, 241, 1195

\bibitem[{{de Freitas Pacheco}(1971)}]{dFP}
{de Freitas Pacheco}, J.~A. 1971, \aap, 13, 58

\bibitem[{{Fermi}(1949)}]{Fermi49}
{Fermi}, E. 1949, Physical Review, 75, 1169

\bibitem[{{Hasselmann} \& {Wibberenz}(1968)}]{HassWib}
{Hasselmann}, K. \& {Wibberenz}, G. 1968, Z. Geophys., 34, 353

\bibitem[{{Hayakawa}(1969)}]{Haya}
{Hayakawa}, S. 1969, {Cosmic ray physics. Nuclear and astrophysical aspects}
  (Interscience Monographs and Texts in Physics and Astronomy, New York:
  Wiley-Interscience, 1969)

\bibitem[{{Hoerandel}(2007)}]{Hoer}
{Hoerandel}, J.~R. 2007, ArXiv Astrophysics e-prints, arXiv:astro-ph/0702370v1

\bibitem[{{Jokipii}(1966)}]{Jokipii}
{Jokipii}, J.~R. 1966, \apj, 146, 480

\bibitem[{{Knapp}(1999)}]{Knapp}
{Knapp}, G.~R. 1999, in Astronomical Society of the Pacific Conference Series,
  Vol. 163, Star Formation in Early Type Galaxies, ed. P.~{Carral} \&
  J.~{Cepa}, 119--+

\bibitem[{{Lerche} \& {Schlickeiser}(1985)}]{LS85}
{Lerche}, I. \& {Schlickeiser}, R. 1985, \aap, 151, 408

\bibitem[{{Lerche} \& {Schlickeiser}(1988)}]{LS88}
{Lerche}, I. \& {Schlickeiser}, R. 1988, \apss, 145, 319

\bibitem[{{Mannheim}(1993)}]{Mannheim}
{Mannheim}, K. 1993, \aap, 269, 67

\bibitem[{{Mannheim} \& {Schlickeiser}(1994)}]{MannSchlicki}
{Mannheim}, K. \& {Schlickeiser}, R. 1994, \aap, 286, 983

\bibitem[{{Owens} \& {Jokipii}(1977)}]{OwJok}
{Owens}, A.~J. \& {Jokipii}, J.~R. 1977, \apj, 215, 677

\bibitem[{{Reimer} {et~al.}(2004){Reimer}, {Protheroe}, \& {Donea}}]{Reimer}
{Reimer}, A., {Protheroe}, R.~J., \& {Donea}, A.-C. 2004, \aap, 419, 89

\bibitem[{{Rieger} {et~al.}(2007){Rieger}, {Bosch-Ramon}, \& {Duffy}}]{Rieger}
{Rieger}, F.~M., {Bosch-Ramon}, V., \& {Duffy}, P. 2007, \apss, 309, 119

\bibitem[{{Schlickeiser}(1983)}]{Schlick83}
{Schlickeiser}, R. 1983, in International Cosmic Ray Conference, Vol.~12,
  International Cosmic Ray Conference, 193--206

\bibitem[{{Schlickeiser}(2002)}]{Schlick1}
{Schlickeiser}, R. 2002, {Cosmic Ray Astrophysics} (Cosmic ray astrophysics /
  Reinhard Schlickeiser, Astronomy and Astrophysics Library; Physics and
  Astronomy Online Library.~Berlin: Springer.~ISBN 3-540-66465-3, 2002)

\bibitem[{{Schlickeiser} {et~al.}(1987){Schlickeiser}, {Sievers}, \&
  {Thiemann}}]{Sievers}
{Schlickeiser}, R., {Sievers}, A., \& {Thiemann}, H. 1987, \aap, 182, 21

\bibitem[{{Skilling}(1975)}]{Skilling}
{Skilling}, J. 1975, \mnras, 172, 557

\bibitem[{{Strong} \& {Moskalenko}(1998)}]{StrMos}
{Strong}, A.~W. \& {Moskalenko}, I.~V. 1998, \apj, 509, 212

\bibitem[{{Sunyaev} \& {Titarchuk}(1980)}]{SunyTit}
{Sunyaev}, R.~A. \& {Titarchuk}, L.~G. 1980, \aap, 86, 121

\bibitem[{{Tavecchio}(2005)}]{Tav}
{Tavecchio}, F. 2005, in The Tenth Marcel Grossmann Meeting. On recent
  developments in theoretical and experimental general relativity, gravitation
  and relativistic field theories, ed. M.~{Novello}, S.~{Perez Bergliaffa}, \&
  R.~{Ruffini}, 512--+

\bibitem[{{Wang} \& {Schlickeiser}(1987)}]{WangSchlick}
{Wang}, Y.-M. \& {Schlickeiser}, R. 1987, \apj, 313, 200

\bibitem[{{Weisstein}(1999)}]{Wolfram}
{Weisstein}, E.~W. 1999, MathWorld -- A Wolfram Web Resource,
  http://mathworld.wolfram.com/ProlateSpheroidalCoordinates.html

\end{thebibliography}

\end{document}